\documentclass[letterpaper]{jpconf}
\usepackage{graphicx}
\begin{document}
\title{Phase Fluctuations near the Chiral Critical Point}

\author{Joseph Kapusta}

\address{School of Physics and Astronomy, University of Minnesota,
Minneapolis, MN 55455, USA }

\ead{kapusta@physics.umn.edu}

\begin{abstract}
The Helmholtz free energy density is parametrized as a function of temperature and baryon density near the chiral critical point of QCD.  The parametrization incorporates the expected critical exponents and amplitudes.  An expansion away from equilibrium states is achieved with Landau theory.  This is used to calculate the probability that the system is found at a density other than the equilibrium one.  Such fluctuations are predicted to be very large in heavy ion collisions.
\end{abstract}

\section{Introduction}

If the up and down quark masses are zero and the strange quark mass is not, the transition between quark-gluon plasma and hadronic matter may be first or second order at zero baryon chemical potential.  If the up and down quark masses are small enough, there may be a phase transition at sufficiently large baryon chemical potential.  This phase transition is predicted to be in the same universality class as liquid-gas phase transitions and the 3D Ising model.  Diverse studies suggest that there is a curve of first-order phase transition in the $\mu$-$T$ plane that terminates in a second-order phase transition at some critical point $(\mu_c,T_c)$.  The location of this chiral critical point has been estimated using various effective field theory models, such as the Namu Jona-Lasinio model \cite{asakawa89}-\cite{scavenius01}, 
a composite operator model \cite{barducci}, a random matrix model \cite{halasz98}, a linear $\sigma$ model \cite{scavenius01}, an effective potential model \cite{hatta02}, and a hadronic bootstrap model \cite{antoniou02}, as well as various implementations of lattice QCD \cite{fodor02}-\cite{gavai05}.  
Reviews have been written by Stephanov \cite{stephanov} and Mohanty \cite{MohantyQM}.  

This subject is of experimental interest because collisions between heavy nuclei at medium to high energy, such as at the future Facility for Antiproton and Ion Research (FAIR), or in future low energy runs at the Relativistic Heavy Ion Collider (RHIC), may provide experimental information on the phase diagram in the vicinity of a critical point.  One characteristic signature would be large fluctuations on an event by event basis \cite{Shuryak}-\cite{nongaussian}.  

The goal of this work \cite{Kapusta} is to understand the equation of state of QCD near the chiral critical point and some of its implications for high energy heavy ion collisions.  Among the requirements are the incorporation of critical exponents and amplitudes and to have sensible limits as temperature $T \rightarrow 0$ and baryon chemical potential $\mu \rightarrow 0$.  This is accomplished by parametrizing the Helmholtz free energy density as a function of $T$ and baryon density $n$ in an appropriate fashion.  Perhaps the closest work that addressed some of these issues blended a parameterization of the 3D Ising model equation of state into quark and hadron equations of state \cite{attract}.  These two parameterizations can perhaps be viewed as alternatives which provide some idea as to the range of uncertainty in how to describe matter near the chiral critical point.

\section{Critical curve}

Numerous studies suggest that the curve in the $T$ vs. $\mu$ plane separating the quark-gluon and hadron phases is approximately quadratic.
\begin{equation}
\left(\frac{T}{T_0}\right)^2 + \left(\frac{\mu}{\mu_0}\right)^2 = 1
\label{critcurve}
\end{equation}  
See figure 1.  The parameters $T_0 = 180$ MeV and $\mu_0 = 1230$ MeV were chosen to match theoretical estimates and experimental data for the transition between the two phases at $\mu =0$ and $T=0$, respectively.  In this example the critical point is chosen to have the temperature $T_c = 140$ MeV.  Along the solid portion of the curve the transition is first order.  It terminates at a second order transition at the critical point.  The dashed portion of the curve indicates that there is no true thermodynamic phase transition separating the phases, only a rapid crossover. 
\begin{figure}[h]
\begin{minipage}{14pc}
\includegraphics[width=14pc,angle=90]{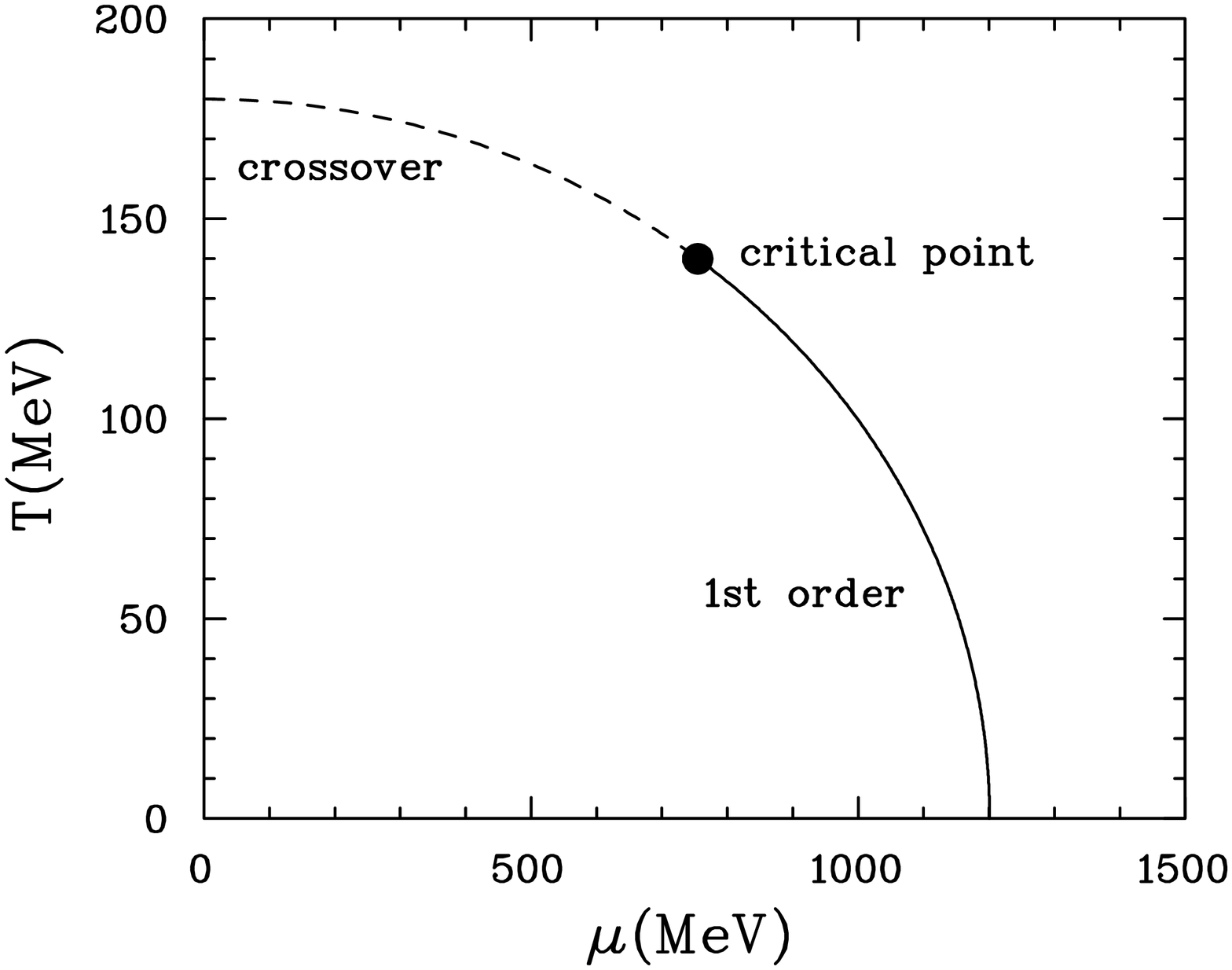}
\caption{ Temperature versus baryon chemical potential from the parameterization given in the text.  The critical point lies somewhere along this curve.}
\end{minipage}
\hspace{6.5pc}%
\begin{minipage}{14pc}
\includegraphics[width=14pc,angle=90]{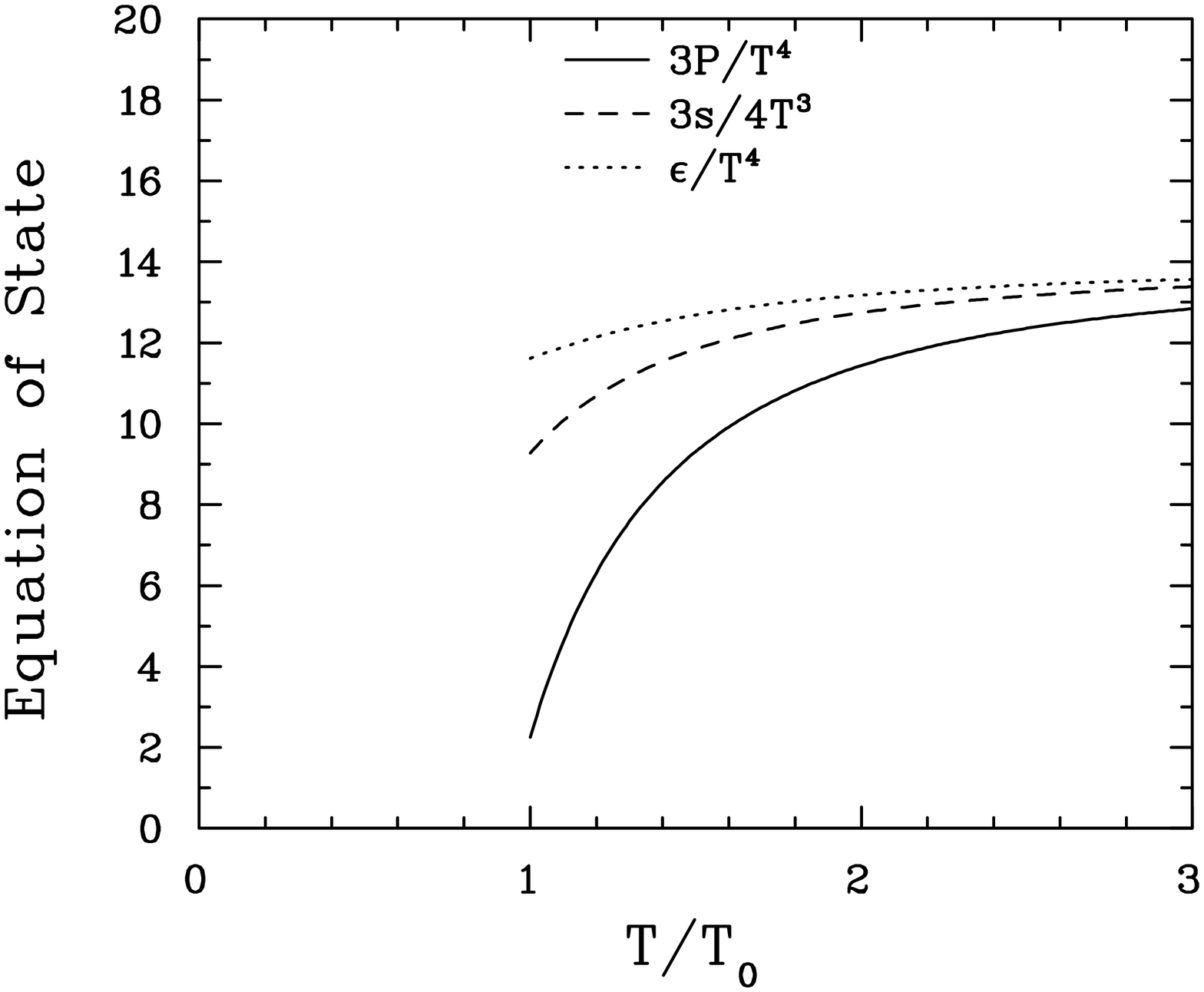}
\caption{ The pressure, entropy density, and energy density, normalized so that they all have the same asymptotic value, at $\mu=0$.  The parameterization is given in the text.}
\end{minipage} 
\end{figure}

We will need the values of the pressure $P_c$, energy density $\epsilon_c$, baryon density $n_c$, and entropy density $s_c$ at the critical point.  The parametrization of the equation of state near the critical point, to be given below, is normalized to these values.  Let us assume that the high energy density equation of state can be parameterized as
\begin{equation}
P = A_4 T^4 + A_2 \mu^2 T^2 + A_0 \mu^4 - B_2 T_0^2 T^2 - B_4 T_0^4 .
\end{equation}
The dimensionless coefficients $A_n$ are adjusted to match a free gas of gluons and 2.5 flavors of massless quarks (the 1/2 to approximately take into account the mass of the strange quark).  The dimensionless coefficients $B_n$ are adjusted to reproduce lattice results near the crossover when $\mu = 0$ \cite{Boyd}-\cite{HotQCD}, and to make the pressure a constant along the critical curve when (\ref{critcurve}) is inserted.  The resulting equation of state for $\mu = 0$ and $T > T_0$ is shown in figure 2.

\section{Equation of state near the critical point}

Given the Helmhotz free energy density $f(n,T)$ the other thermodynamic quantites follow uniquely from it, as summarized below.
\begin{eqnarray}
P &=& n^2 \frac{\partial}{\partial n} \left( \frac{f(n,T)}{n}\right)\\
\mu &=& \frac{\partial f(n,T)}{ \partial n}\\
s &=& - \frac{\partial f(n,T)}{ \partial T}\\
\epsilon &=& f(n,T) + Ts(n,T)
\end{eqnarray}
These satisfy the thermodynamic identity $\epsilon = -P +Ts + \mu n$.  It is useful to define the dimensionless variables $t = (T - T_c)/T_c$ and $\eta = (n - n_c)/n_c$ which are a measure of distance from the critical point in temperature and density.  Near the critical point the heat capacity $c_V$ has the divergent behavior
\begin{eqnarray}
c_V = T \frac{\partial s(n,T)}{\partial T} \rightarrow
 \left\{ \begin{array}{ll}
c_- (-t)^{-\alpha} & \mbox{when $t \rightarrow 0^-$} \\
c_+ t^{-\alpha} & \mbox{when $t \rightarrow 0^+$ ,}
\end{array} \right.
\end{eqnarray}
while the thermal conductivity $\kappa_T$ and baryon susceptibility $\chi_B$ have the divergent behavior
\begin{eqnarray}
\frac{\chi_B}{n^2} = \kappa_T = \left[ n \frac{\partial P(n,T)}{\partial n} \right]^{-1} \rightarrow
 \left\{ \begin{array}{ll}
\kappa_- (-t)^{-\gamma} & \mbox{when $t \rightarrow 0^-$} \\
\kappa_+ t^{-\gamma} & \mbox{when $t \rightarrow 0^+$ .}
\end{array} \right.
\end{eqnarray}
Along the coexistence curve the density difference between the higher density liquid phase (quarks and gluons) and the lower density gas phase (hadrons) goes to zero as
\begin{equation}
n_l - n_g \sim (-t)^\beta .
\end{equation}
Along the critical isotherm the pressure behaves as
\begin{equation}
P - P_c \sim |\eta|^{\delta} \, {\rm sign}(\eta) .
\end{equation}
Here $\alpha$, $\beta$, $\gamma$ and $\delta$ are critical exponents.  These exponents are related by $\alpha + 2\beta + \gamma = 2$ and $\gamma = \beta (\delta - 1)$.  Mean field theories normally give $\alpha = 0$, $\beta = 1/2$, $\gamma = 1$, and $\delta = 3$.  Typical fluids are measured to have $\alpha \ll 1$, $\beta \approx 1/3$, $1.2 < \gamma < 1.3$, and $4 < \delta < 5$ \cite{fluidexpts}.  The 3D Ising model has $\alpha = 0.11$, $\beta = 0.325$, $\gamma = 1.24$, and $\delta = 4.815$ \cite{Guida}, which are the values used here.  Furthermore, the critical amplitudes are related by $\kappa_+/\kappa_- \approx 5$ and $c_+/c_- \approx 0.5$, which is universal to all theories within the same class.  See \cite{Zinnbook}.

The challenge now is to parametrize $f(\eta,t)$ so that it yields the correct critical behavior near $\eta = t = 0$ and still gives sensible results away from the critical point.  Mean field theories normally result in a Taylor series expansion in integral powers of $\eta$ with coefficients that depend on $t$.  The simplest way to incorporate the correct critical behavior is with a modified expansion of the form
\begin{equation}
f = f_0(t) + f_1(t)\eta + f_2(t)\eta^2 + f_{\sigma}(t)|\eta|^{\sigma} . 
\end{equation}
The power $\sigma = \delta + 1 = 5.815$.  Higher powers of $\eta$ could be included but they would not affect the critical behavior and so are discarded here for simplicity.  The coefficient functions must have the following behavior.
\begin{eqnarray}
f_0(t) = \left\{ \begin{array}{ll}
\bar{f}_0(t) -a_- (-t)^{2-\alpha} & \mbox{if $t<0$} \\
\bar{f}_0(t) -a _+ t^{2-\alpha} & \mbox{if $t>0$ .}
\end{array} \right.
\end{eqnarray}
\begin{equation}
f_1(t) = n_c \mu_0 \sqrt{1 - \frac{T_c^2}{T_0^2}(1+t)^2 } .
\end{equation}
\begin{eqnarray}
f_2(t) = \left\{ \begin{array}{ll}
\bar{f}_2(t)-b_- (-t)^{\gamma} & \mbox{if $t<0$}\\
\; \;\; \bar{f}_2(t) + b_+ t^{\gamma} & \mbox{if $t>0$}
\end{array} \right.
\end{eqnarray}
Here $\bar{f}_0(t)$ and $\bar{f}_2(t)$ are smooth functions of $t$ with $\bar{f}_2(0) = 0$.  They are parametrized so that $f(n,T)$ has sensible limits at $T=0$ and at $n=0$.  The coefficient $f_{\sigma}$ is assumed to be constant and proportional to $P_c$; for definiteness we shall take $f_{\sigma} = 5P_c \approx 512$ MeV/fm$^3$.  Results for the critical curves, latent heat, thermal conductivity, and heat capacity are shown in figures 3 to 6.

\begin{figure}[h]
\begin{minipage}{14pc}
\includegraphics[width=14pc,angle=90]{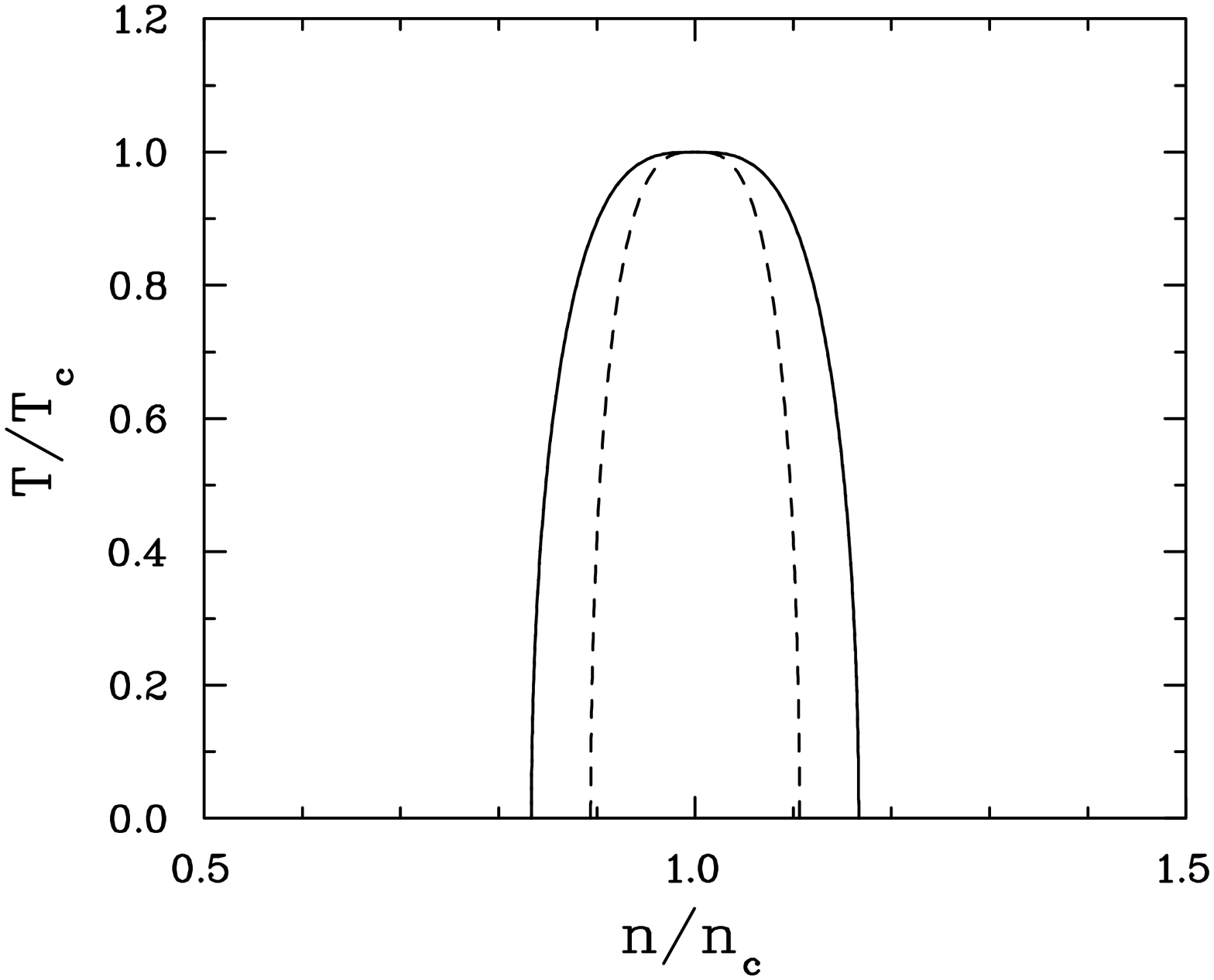}
\caption{The solid curve denotes coexistence between high and low density phases and the dashed curve denotes the limits of metastability.  The scaled curves are independent of the choice of critical temperature and density.}
\end{minipage}
\hspace{6.5pc}%
\begin{minipage}{14pc}
\includegraphics[width=14pc,angle=90]{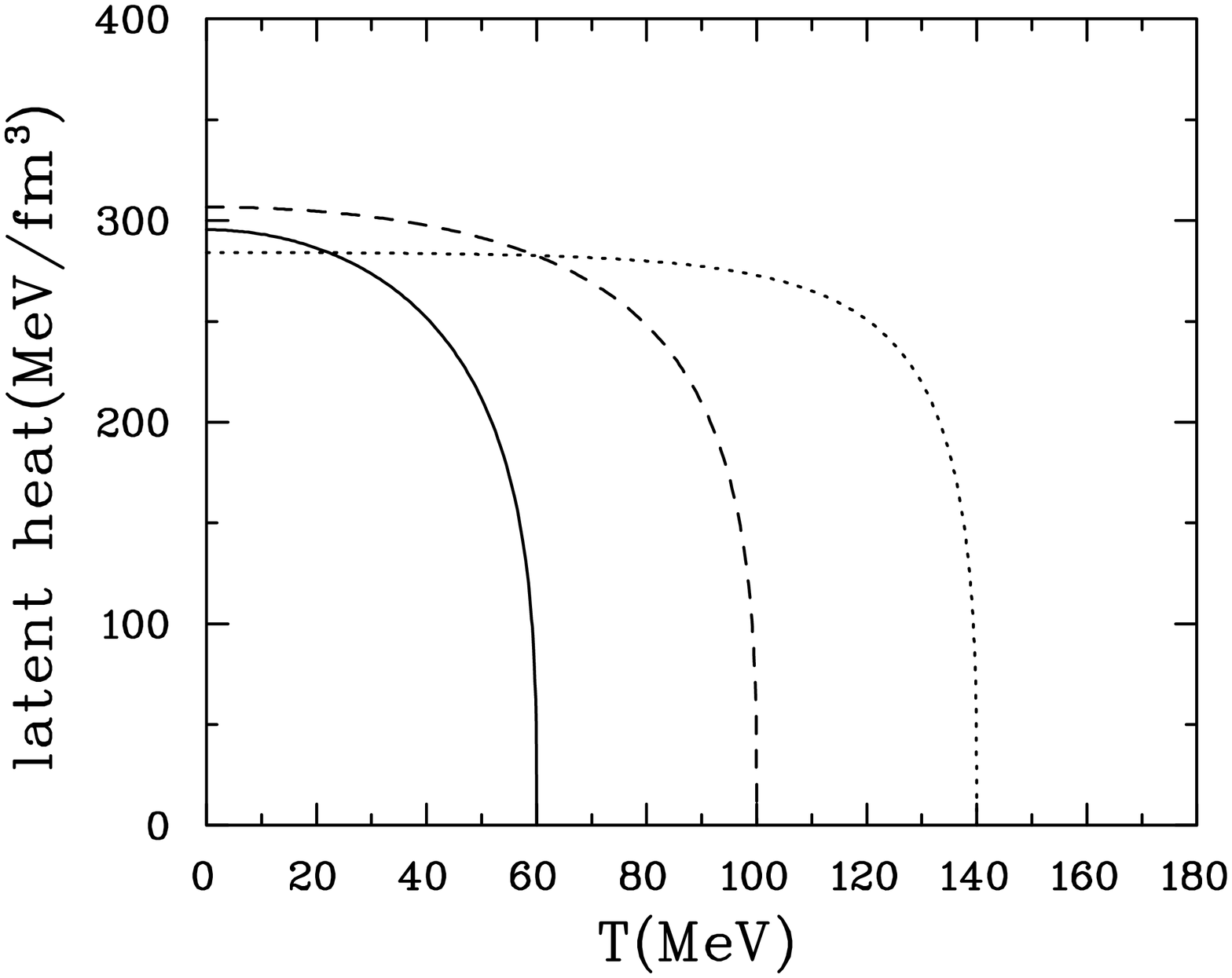}
\caption{The latent heat per unit volume versus temperature for three choices of critical temperature.}
\end{minipage} 
\end{figure}

\begin{figure}[h]
\begin{minipage}{14pc}
\includegraphics[width=14pc,angle=90]{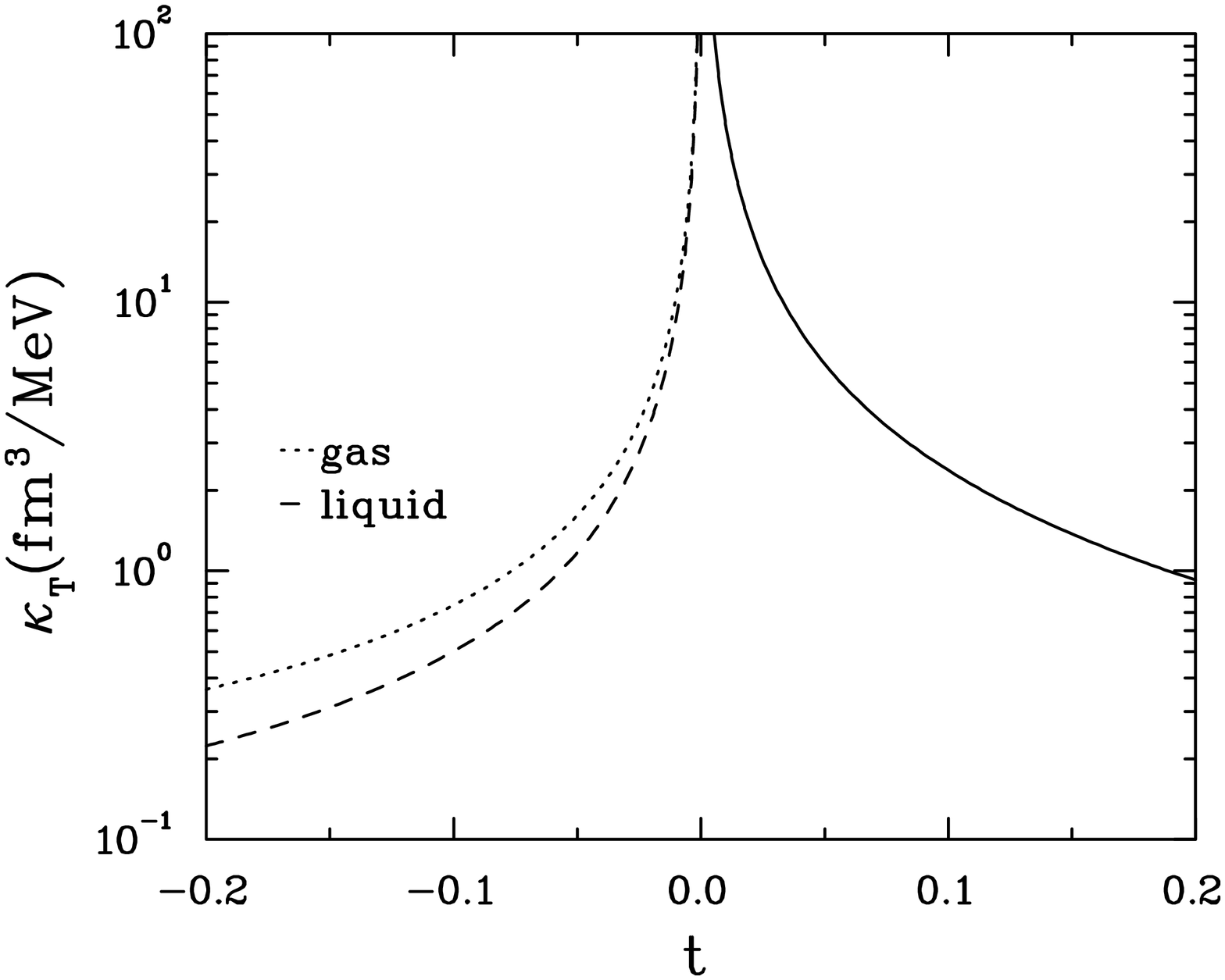}
\caption{The isothermal compressibility.  For $t<0$ they are evaluated along the coexistence curve while for $t>0$ it is evaluated at the critical density.  The curves for different critical temperature fall on top of one another.}
\end{minipage}
\hspace{6.5pc}%
\begin{minipage}{14pc}
\includegraphics[width=14pc,angle=90]{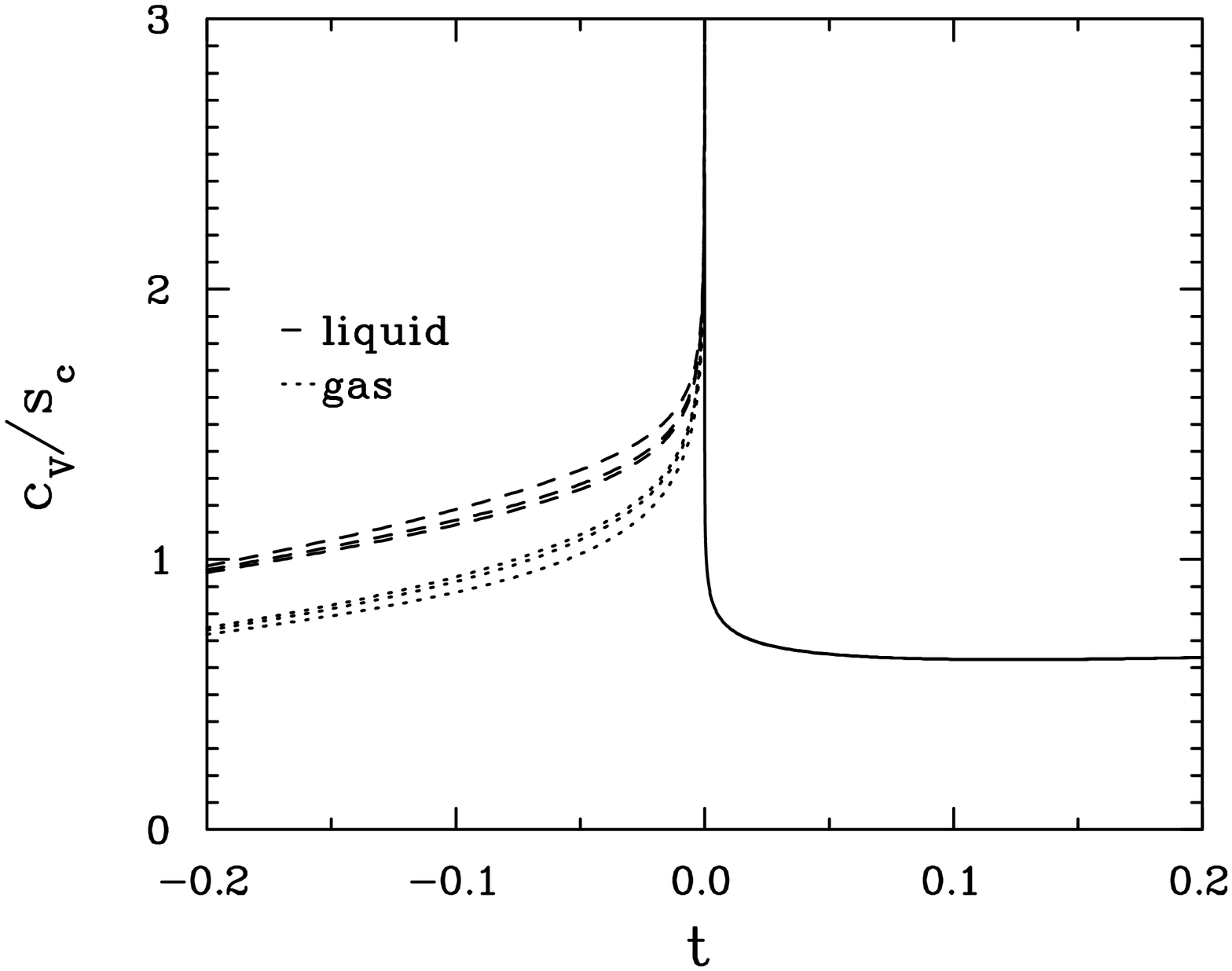}
\caption{The heat capacity per unit volume.  For $t<0$ they are evaluated along the coexistence curve while for $t>0$ it is evaluated at the critical density.  When divided by the entropy density at the critical point the results are nearly independent of the choice of critical temperature.}
\end{minipage} 
\end{figure}

\section{Fluctuations}

In any system of finite size there will be thermal fluctuations.  These fluctuations become large as the critical point is approached.  Fluctuations can be studied with Landau theory \cite{Landau,Goodman}.  The question is: What is the probability to find a system of volume $V$ with a baryon density different from the equilibrium one?  The expansion away from equilibrium states is determined by the thermodynamic potential
\begin{equation}
\Omega(\mu,T;\eta) - \Omega_0(\mu,T) = \left[ (f_1 -n_c \mu) \eta
+ f_2 \eta^2 + f_{\sigma}|\eta|^{\sigma} \right] V .
\end{equation}
Here $\Omega_0(\mu,T) = (f_0 - n_c \mu)V$. Along the coexistence curve $f_1 = n_c \mu$.  The relative probability to be at a density other than the equilibrium one, along the coexistence curve, is
\begin{equation}
{\cal P}(\eta)/{\cal P}(\eta_l) = \exp\left(-\Delta \Omega/T\right)
\end{equation}
where
\begin{equation}
\Delta \Omega = \left[ f_2 \left( \eta^2 - \eta_l^2 \right)
+ f_{\sigma} \left( |\eta|^{\sigma} - |\eta_l|^{\sigma} \right) \right] V .
\end{equation}
For purposes of illustration the volume is taken to be 400 fm$^3$.  The value of 400 fm$^3$ is about as large as one can imagine for high energy nuclear collisions.  Considering that the critical density is estimated to be about $5n_0 \approx 0.75$ baryons/fm$^3$, this would mean that about 300 baryons participate in the fluctuation.  That is a substantial fraction of the total in Au+AU, Pb+Pb, or in U+U collisions.  Smaller volumes would have even larger fluctuations.

\begin{figure}[h]
\begin{minipage}{14pc}
\includegraphics[width=14pc,angle=90]{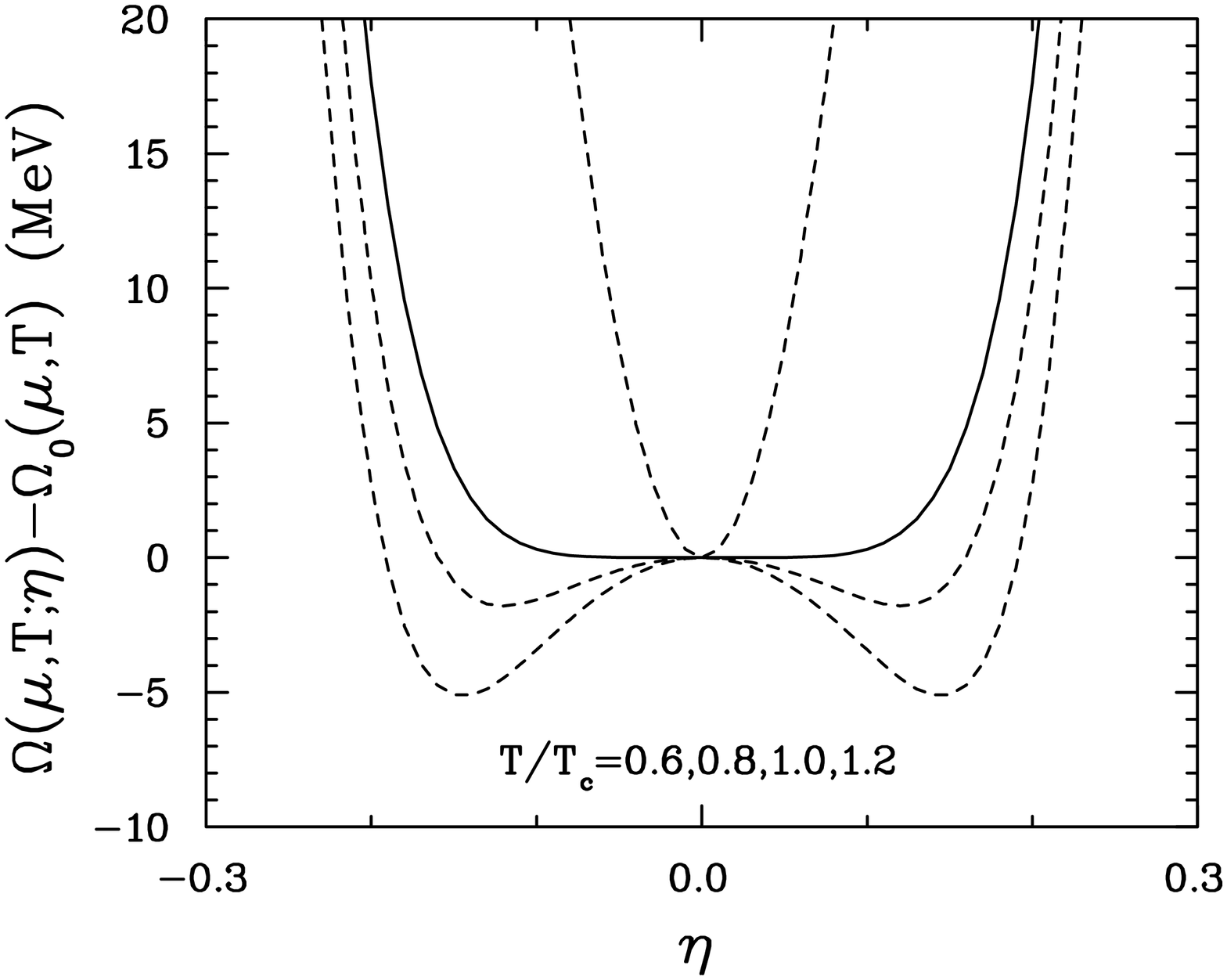}
\caption{Thermodynamic potential as a function of $\eta$ near the critical point when the volume is 400 fm$^3$.  The stable phases are located at the minima of the potential.  Four different temperatures are shown, with the solid curve representing the critical temperature.}
\end{minipage}
\hspace{6.5pc}%
\begin{minipage}{14pc}
\includegraphics[width=14pc,angle=90]{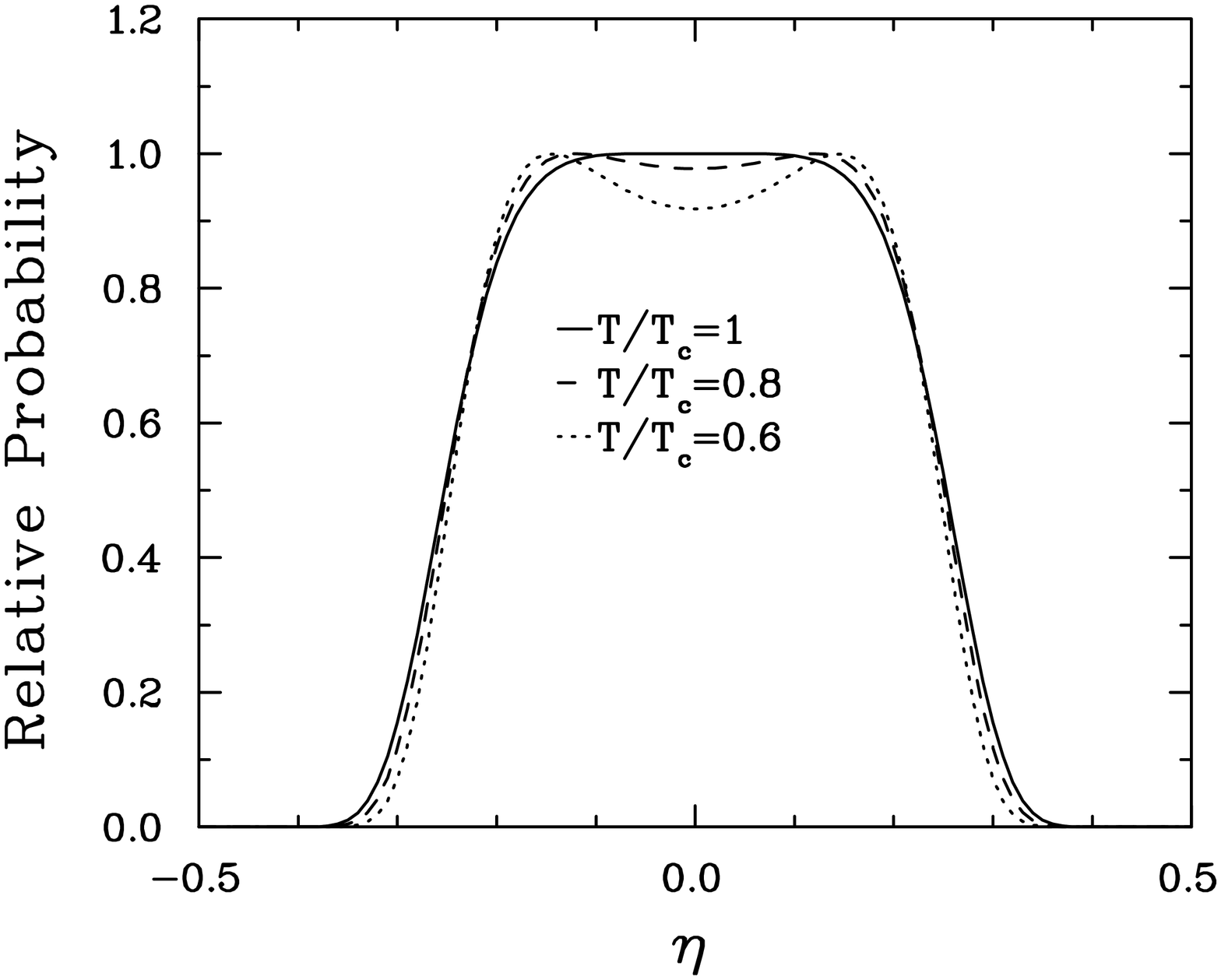}
\caption{The probability to find the system at a particular density relative to the equilibrium densities at phase coexistence.}
\end{minipage} 
\end{figure}

The thermodynamic potential is shown in figure 7 for several different temperatures.  Phase coexistence occurs below $T_c$; the two phases are located at the minima of the potential.  They are not very deep, indicating that fluctuations do not cost much free energy.  The potential is so flat because $f_2(T_c)=0$ and because the power $\sigma \approx 6$ is so large.  The shallowness of the potential is reflected in the probabilities, as shown in figure 8.  The probability to find the system with a density anywhere between $\pm 25$\% of the critical density is greater than 50\%.  Fluctuations are large!  They would be even larger for smaller volume systems.

\section{Conclusions}

An equation of state valid in the vicinity of the chiral critical point has been constructed.  It incorporates correct values of the critical exponents and amplitudes.  Since only certain properties of the equation of state are universal, there is some freedom to vary the noncritical functional dependence on temperature and density.  Work on extending the equation of state to a wider range of $T$ and $\mu$ is underway.

The Landau theory of fluctuations away from equilibrium states was used to determine the magnitude of the fluctuations one might expect in heavy ion collisions.  These fluctuations are quite large, partly due to finite volume effects but mostly because the critical exponent $\delta$ is much larger than in mean field theories.  This flattens the Landau free energy as a function of density away from the equilibrium densities and so decreases the cost to fluctuate away from them.

It will require careful thought as to how to incorporate fluctuations near the critical point into dynamical simulations of heavy ion collisions.  What is the appropriate way to describe the transition in heavy ion collisions?  Is it nucleation \cite{nucleate}, spinodal decomposition \cite{Randrup}, or something else?  What are the best experimental observables and can they be measured at RHIC and/or FAIR?  The future of this topic is exciting!

\section*{Acknowledgements}

This work was supported by the US Department of Energy (DOE) under Grant No. 
DE-FG02-87ER40328.

\end{document}